\documentclass[pra, twocolumn]{revtex4}
\usepackage{graphicx, amsmath, amssymb, bm}

\begin{document}

\title{Supersensitive measurement of angular displacements using entangled photons}

\author{Anand Kumar Jha$^1$}
\email{akjha9@gmail.com}

\author{Girish S. Agarwal$^2$}
\email{girish.agarwal@okstate.edu}

\author{Robert W. Boyd$^1$}
\email{boydrw@mac.com}

\affiliation{$^1$The Institute of Optics, University of Rochester,
Rochester, New York 14627, USA \\
$^2$Department of Physics, Oklahoma State University, Stillwater,
Oklahoma 74078, USA}

\date{\today}

\begin{abstract}

We show that the use of entangled photons having non-zero orbital
angular momentum (OAM) increases the resolution and sensitivity of
angular-displacement measurements performed using an
interferometer. By employing a 4$\times$4 matrix formulation to
study the propagation of entangled OAM modes, we analyze
measurement schemes for two and four entangled photons and obtain
explicit expressions for the resolution and sensitivity in these
schemes. We find that the resolution of angular-displacement
measurements scales as $Nl$ while the angular sensitivity
increases as $1/(2Nl)$, where $N$ is the number of entangled
photons and $l$ the magnitude of the orbital-angular-momentum mode
index. These results are an improvement over what could be
obtained with $N$ non-entangled photons carrying an orbital
angular momentum of $l\hbar$ per photon.

\end{abstract}

\maketitle

\section{introduction}

Precision measurements are important not only for verifying a
given physical theory but also for possible applications of the
theory. For example, the fact that relative displacements can be
measured with sub-wavelength sensitivity through optical phase
measurements has led to many useful applications in a wide variety
of fields including cosmology, nanotechnology, metrology and
medicine.

In generic classical schemes for optical phase measurements, the
sensitivity is limited by what is known as the standard quantum
limit, which scales as $1/\sqrt{N}$, where $N$ is either the
average number of photons in the coherent state input to the
interferometer or the number of times the experiment is repeated
with one-photon fock-state input \cite{scully&zubairy,
lee2002jmo}. More recent works have shown that the use of
non-classical states of light can lead to improved sensitivity in
optical phase measurements \cite{caves1981prd, yurke1986prl,
boto2000prl, dangelo2001prl}. In particular, it has been shown
that an $N$-photon entangled-state input to an interferometer
gives rise to phase super-resolution \cite{boto2000prl,
walther2004nature, mitchell2004nature, kolkiran2007optexp,
sun2006pra}, that is, the narrowing of interference fringes by $N$
times compared to the fringes obtained with classical schemes at
the same wavelength. It has also been shown that with $N$
entangled photons the sensitivity of optical phase measurements
scales as $1/N$, in contrast to the $1/\sqrt{N}$ scaling obtained
using $N$ non-entangled photons \cite{nagata2007science,
resch2007prl}. The $1/N$ scaling is also known as the Heisenberg
limit.

In this paper, we consider an analogous type of measurement,
namely, angular-displacement measurements.  We seek to determine
how accurately the angular orientation of an optical component can
be measured using purely optical methods.  Specifically, we
consider an optical component in the form of a Dove prism and seek
to measure its angular orientation by determining the rotation
angle induced in an optical beam in passing through the prism.  We
assume that the prism is located in one arm of an interferometer.
We thus seek to answer the question as to how accurately the
angular displacements (rotations) introduced in a beam of light
inside an interferometer can be measured.  Measurements of this
sort are generic to a broad class of problems in quantum
metrology. We explicitly analyze measurement schemes for two and
four entangled photons and compare the angular resolution and
sensitivity with those obtained using classical measurement
schemes. We find that the use of entangled photons with non-zero
orbital angular momentum increases the resolution and sensitivity
of angular displacement measurements. The angular resolution
increases by a factor of $Nl$ while the angular sensitivity
increases as $1/2Nl$, where $N$ is the number of entangled photons
and $l$ the magnitude of the orbital-angular-momentum mode index.

The paper is organized as follows. In Sec. II, we obtain analytic
expressions for the two- and four-photon fields produced by
parametric down-conversion. In Sec. III, we employ a $4\times4$
matrix formulation to study the propagation of entangled OAM modes
through various optical elements and illustrate super-sensitive
angular-displacement measurements with two and four entangled
photons. Section IV presents our conclusions.

\section{Entangled photons produced by parametric
down-conversion}

We start with the following interaction Hamiltonian $\hat{H}(t)$
for parametric down-conversion (PDC) \cite{ou1989pra}:
\begin{multline}
\hat{H}(t)=\frac{\epsilon_0}{2}\int_{\mathcal{V}} d^{3}\bm{r}
\chi^{(2)} E_{0}^{(+)}(\bm{r},t) \hat{E_{s}}^{(-)}(\bm{r},t)
\hat{E_{i}}^{(-)}(\bm{r},t) \\ + {\rm H.c.}, \label{interaction
hamiltoinan1}
\end{multline}
where $\mathcal{V}$ is the volume of the interacting part of the
nonlinear crystal and $\chi^{(2)}$ is the second-order nonlinear
susceptibility. $\hat{E_{j}}^{(+)}(\bm{r},t)$ and
$\hat{E_{j}}^{(-)}(\bm{r},t)$ are the positive- and
negative-frequency parts of the electric field, where $j=s$ and
$i$ stand for the signal and idler, respectively. The pump field
$E_0$ is assumed to be strong and will therefore be treated
classically. We decompose the three electric fields in terms of
field-modes $u_{p}^{l}(\bm r)$ carrying orbital angular momentum
(OAM). These modes are characterized by two indices, $l$ and $p$,
and carry an OAM of $l\hbar$ per photon owing to their azimuthal
phase dependence of $e^{il\phi}$ \cite{allen1992pra}. The index
$l$ is referred to as the OAM mode index. The modes $u_{p}^{l}(\bm
r)$ are assumed to have the following general form
\begin{align}
u_{p}^{l}(\bm r)=R_p(\rho,z)\frac{e^{il\phi}}{\sqrt{2\pi}},
\label{mode-function}
\end{align}
with $R_p(\rho, z)$ being a complete set of orthonormal, radial
modes, that is, $\sum_p \rho R_p(\rho,z)^{*}R_p(\rho',z)=\delta
(\rho-\rho')$ and $\int \rho d\rho
R_p(\rho,z)^{*}R_{p'}(\rho,z)=\delta_{pp'}$. One possible choice
for $u_{p}^{l}(\bm r)$ are the Laguerre-Gaussian modes, but the
best choice is usually the Schimdt modes of the down-converted
field, which, in general, are not the Laguerre-Gaussian modes
\cite{law2004prl}. The three electric fields can now be written as
\begin{align}
&E_{0}^{(+)}(\bm r, t)=A(\omega_0)  u_{p_0}^{l_0}(\bm r)
e^{-i\omega_{0}
t},\label{quantize-pump} \\
&\hat{E_{s}}^{(-)}(\bm r,
t)=\sum_{l_s,p_s}\hat{a}_{l_s,p_s}^{\dag}(\omega_{s})u^{l_s
*}_{p_s}(\bm
r)e^{i\omega_{s} t}\label{quantize-signal}, \\
&\hat{E_{i}}^{(-)}(\bm r,
t)=\sum_{l_i,p_i}\hat{a}_{l_i,p_i}^{\dag}(\omega_{i})u^{l_i
*}_{p_i}(\bm r) e^{i\omega_{i} t}\label{quantize-idler}.
\end{align}
Here we have assumed that the signal, idler and pump fields are
monochromatic with frequencies $\omega_s$, $\omega_i$ and
$\omega_0$, respectively. The three fields interact for some time
within the nonlinear crystal and the state $|\psi\rangle $ of the
down-converted photons after the interaction is given by $
|\psi\rangle =\mathcal{T}\left\{\exp\left[1/(i\hbar) \int dt
\hat{H}(t)
  \right]\right\}|\psi_0\rangle$,
where
$|\psi_0\rangle=|\textrm{vac}\rangle_{s}|\textrm{vac}\rangle_{i}$
is the initial vacuum state before the interaction, with no
photons in either the signal or the idler mode. We assume perfect
frequency phase-matching such that $\omega_0=\omega_s+\omega_i$;
the symbol $\mathcal{T}$ represents operator time-ordering. Taking
the parametric interaction to be very weak, we then write the
state $|\psi\rangle $ in terms of a perturbative expansion
\cite{ou1989pra}:
\begin{align}\label{temp-two-photon-state}
|\psi\rangle = |\psi_0\rangle + |\psi_2\rangle+ |\psi_4\rangle +
\cdots
\end{align}
The first term $|\psi_0\rangle$ is the initial vacuum state, the
second term $|\psi_2\rangle\equiv\mathcal{T}\left[1/(i\hbar)\int
dt \hat{H}(t)\right]|\psi_0\rangle$ is the two-photon state and
the third term $|\psi_4\rangle\equiv
\mathcal{T}\left[-1/(2\hbar^2) \iint dtdt'
\hat{H}(t)\hat{H}(t')\right]|\psi_0\rangle$ is the four-photon
state, etc.

We calculate the two-photon state $|\psi_2\rangle$ by substituting
from Eqs.~(\ref{interaction hamiltoinan1}), (\ref{quantize-pump}),
(\ref{quantize-signal}) and (\ref{quantize-idler}) into
Eq.~(\ref{temp-two-photon-state}) and obtain
\begin{multline}
|\psi_2\rangle =\Big[\frac{\epsilon_0 \chi^{(2)}
A(\omega_0)}{2i\hbar}\sum_{l_s,p_s}\sum_{l_i,p_i}\hat{a}_{l_s,p_s}^{\dag}
\hat{a}_{l_i,p_i}^{\dag} \\ \times \int_{\mathcal{V}} d^{3}\bm{r}
u^{l_s *}_{p_s}(\bm r) u^{l_i *}_{p_i}(\bm r)u_{p_0}^{l_0}(\bm r)
+ {\rm H.c.} \Big] |{\rm vac}\rangle_s|{\rm
  vac}\rangle_i.
\end{multline}
Working in the cylindrical coordinate system and using the
orthogonality relation $\int_0^{2\pi} d\phi
e^{i(l_0-l_s-l_i)\phi}=2\pi\delta_{l_0, l_s+l_i}$, we arrive at
the following expression for the two-photon state:
\begin{align}
|\psi_2\rangle
=\sum_{l,p_s,p_i}\chi_{l,p_s,p_i}|l,p_s\rangle_s|-l,p_i\rangle_i,
\end{align}
where
\begin{multline}
\chi_{l,p_s,p_i}=\frac{\epsilon_0 \chi^{(2)}
A(\omega_0)}{2i\hbar}\\ \times \iint \rho d\rho dz
R_{p_s}(\rho,z)R_{p_i}(\rho,z)R_{p_0}(\rho, z)
\end{multline}
is the probability amplitude that the signal and idler photons are
in the modes characterized by indices $(l, p_s)$ and $(-l, p_i)$,
respectively. Next, we consider a detection system that is
insensitive to the radial indices and is sensitive only to the OAM
mode index. With respect to such detection systems the above state
can be written as
\begin{align}
|\psi_2\rangle =\sum_{l}\sqrt{P_l}|l\rangle_s|-l\rangle_i,
\label{two-photon-state1}
\end{align}
where
\begin{align}
P_l=\sum_{p_s,p_i}|\chi_{l,p_s,p_i}|^2
\end{align}
is the probability that the orbital angular momenta of the signal
and idler photons are $l\hbar$ and $-l\hbar$, respectively.

The next term in the expansion of
Eq.~(\ref{temp-two-photon-state}) is the four-photon state
$|\psi_4\rangle$. We evaluate this term in a similar manner, and,
assuming a detection system sensitive only to the OAM mode index,
we obtain
\begin{align}
|\psi_4\rangle
=\sum_{l,l'}\sqrt{P_{l,l'}}|l,l'\rangle_s|-l,-l'\rangle_i,
\label{four-photon-state1}
\end{align}
where
\begin{align}
P_{l,l'}=\sum_{p_s,p_i}\sum_{p'_s,p'_i}|\chi_{l,p_s,p_i}|^2|\chi_{l,p'_s,p'_i}|^2
\end{align}
is now the probability that two photons with OAMs $l\hbar$ and
$l'\hbar$ are produced in the signal mode and two photons with
OAMs $-l\hbar$ and $-l'\hbar$ are produced in the idler mode.

\section{Supersensitive measurement of angular displacements}

In this section we describe our measurement schemes for
supersensitive angular-displacement measurements. Our proposed
schemes are very analogous to those proposed by Kolkiran and
Agarwal in the context of phase supersensitivity
\cite{kolkiran2007optexp}.

First, we make some notational changes from the previous section
to render the subsequent calculations less cumbersome. From this
section onwards, we use the mode label to also represent the
annihilation operator corresponding to that mode. For example,
$\hat{s}$ and $\hat{i}$ represent the annihilation operators
corresponding to the signal($s$) and idler($i$) modes,
respectively. Further, a given mode is separated into two
different modes, one corresponding to the positive value of the
orbital angular momentum and the other corresponding to the
negative value. Thus $\hat{s}_{-l}$ represents the annihilation
operator corresponding to the signal mode having an OAM of
$-l\hbar$, etc. In accordance with the above notation, we also
change the notation for representing the state of the photons. For
example, $|2\rangle_{s_{+l}}$ represents two photons in $s$ mode
with orbital angular momentum $+l\hbar$ per photon.

\begin{figure}
\centering
\includegraphics{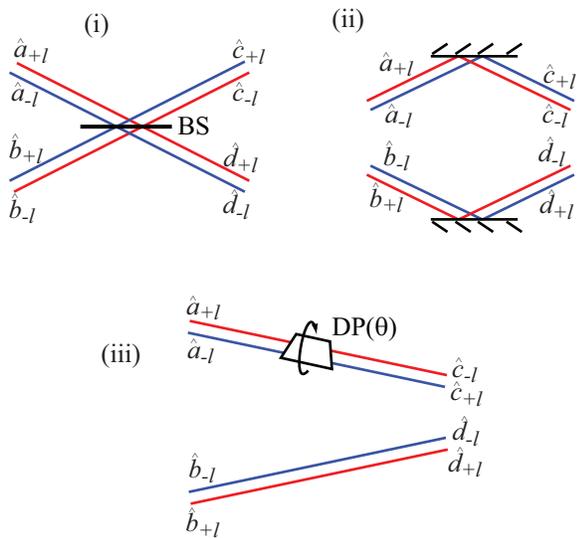}
\caption{ Transformation of OAM modes when they pass through (i) a
beam splitter, (ii) a pair of mirrors, and (iii) a Dove prism.}
\label{fig0}
\end{figure}

Next, we summarize the transformation properties of OAM modes when
they pass through either a beam splitter(BS), a pair of mirrors or
a Dove prism. We note that upon reflection an OAM mode changes the
sign of its mode index and also picks up an additional phase. This
additional phase is equal to $\pi/2$ when the mode reflects from a
symmetric beam splitter and is equal to $\pi$ when it reflects
from a mirror \cite{gonzalez2006optexp}. The beam splitter
transformation matrix is calculated in the following way. As shown
in Fig.~\ref{fig0}(i), let us suppose that $a$ and $b$ are the
input modes to a beam splitter and $c$ and $d$ are the output
modes. The annihilation operators corresponding to mode $a$ are
$\hat{a}_{+l}$ and $\hat{a}_{-l}$, etc. Using the standard
beam-splitter operator algebra \cite{mandel&wolf}, we obtain the
relationship between the input and output mode annihilation
operators and represent it as the matrix equality:
\begin{align}
\left(%
\begin{array}{c}
  \hat{c}_{+l} \\
  \hat{c}_{-l} \\
  \hat{d}_{+l} \\
  \hat{d}_{-l} \\
\end{array}%
\right)=\frac{1}{\sqrt{2}}
\left(%
\begin{array}{cccc}
  0 & i & 1 & 0 \\
  i & 0 & 0 & 1 \\
  1 & 0 & 0 & i \\
  0 & 1 & i & 0 \\
\end{array}%
\right)\left(%
\begin{array}{c}
  \hat{a}_{+l} \\
  \hat{a}_{-l} \\
  \hat{b}_{+l} \\
  \hat{b}_{-l} \\
\end{array}%
\right)=M_{\rm bs}\left(%
\begin{array}{c}
  \hat{a}_{+l} \\
  \hat{a}_{-l} \\
  \hat{b}_{+l} \\
  \hat{b}_{-l} \\
\end{array}%
\right). \label{matrix-bs}
\end{align}
Here the unitary matrix $M_{\rm bs}$ is the beam splitter
transformation matrix for OAM modes. In a similar manner, the
transformation matrix $M_{\rm mir}$ related to the reflections of
two incident modes $a$ and $b$ into the reflected mode $c$ and $d$
[Fig.~\ref{fig0}(ii)] can be shown to be
\begin{align}
M_{\rm mir}=\left(%
\begin{array}{cccc}
  0 & -1 & 0 & 0 \\
  -1 & 0 & 0 & 0 \\
  0 & 0 & 0 & -1 \\
  0 & 0 & -1 & 0 \\
\end{array}%
\right). \label{matrix-mirror}
\end{align}
Finally, in situations in which one of the modes passes through a
Dove prism [Fig.~\ref{fig0}(iii)], rotated at an angle $\theta$,
the transformation matrix $M_{\rm dp}$ is given by
\begin{align}
M_{\rm dp}(\theta)=\left(%
\begin{array}{cccc}
  0 & -e^{2il\theta} & 0 & 0 \\
  -e^{-2il\theta} & 0 & 0 & 0 \\
  0 & 0 & 1 & 0 \\
  0 & 0 & 0 & 1 \\
\end{array}%
\right). \label{matrix-dp}
\end{align}
The two non-zero off-diagonal matrix elements show that, upon
passage through a Dove prism, an OAM mode picks up an additional
phase of $\pi-2l\theta$ \cite{gonzalez2006optexp}, where $l$ is
the orbital angular momentum mode index and $\theta$ is the angle
of rotation of the Dove prism. The first two diagonal elements are
zero due to the fact that upon passage through a Dove prism a
modes changes the sign of its OAM mode index. We note that both
$M_{\rm mir}$ and $M_{\rm bs}$ are unitary matrices.

We are now ready to analyze the situations shown in
Figs.~\ref{fig1} and \ref{fig2}. In these schemes, the entangled
photons produced by PDC in modes $s$ and $i$ are mixed at a beam
splitter (BS1). A $\theta$ rotation of the Dove prism (DP) then
rotates the photon-field in mode $g$ with respect to the field in
mode $h$. The photons are then mixed at the second beam splitter
(BS2) and detected subsequently in modes $a$ and $b$. Our aim is
to determine the resolution and sensitivity with which the
rotation angle $\theta$ can be measured. Using the transformation
properties of OAM modes as given by Eqs.~(\ref{matrix-bs}),
(\ref{matrix-mirror}) and (\ref{matrix-dp}), we express the output
mode annihilation operators in terms of the input mode
annihilation operators as
\begin{align}
O&=M_{\rm bs}M_{\rm dp}(\theta)M_{\rm mir}M_{\rm bs}M_{\rm mir}I=M
I, \label{matrix-total}
\end{align}
where
\begin{align}
O=\left(%
\begin{array}{c}
  \hat{a}_{+l} \\
  \hat{a}_{-l} \\
  \hat{b}_{+l} \\
  \hat{b}_{-l} \\
\end{array}%
\right) \quad {\rm and} \quad  I=\left(%
\begin{array}{c}
  \hat{s}_{+l} \\
  \hat{s}_{-l} \\
  \hat{i}_{+l} \\
  \hat{i}_{-l} \\
\end{array}%
\right). \notag
\end{align}
In order to calculate the state in the output modes $a$ and $b$,
we need to obtain the inverse relationship, that is, we need to
express the input mode annihilation operators in terms of the
output mode annihilation operators. Therefore, we invert the above
matrix equation and write it as
\begin{align}
I=M^{-1}O=M^{\dagger}O,
\end{align}
where the last equality results from the fact that $M$ is a
unitary matrix $(M^{-1}=M^{\dagger})$, with $|{\rm det} M|=1$,
where det denotes the determinant. Now taking the transpose of the
above equation we obtain:
\begin{align}
I^{\dagger}=O^{\dagger}M. \label{input-output}
\end{align}
We note that $I^{\dagger}$ and $O^{\dagger}$ are four-element row
vectors: $I^{\dagger}=(\hat{s}^{\dagger}_{+l},
\hat{s}^{\dagger}_{-l}, \hat{i}^{\dagger}_{+l},
\hat{i}^{\dagger}_{-l})$ and $O^{\dagger}=(\hat{a}^{\dagger}_{+l},
\hat{a}^{\dagger}_{-l}, \hat{b}^{\dagger}_{+l},
\hat{b}^{\dagger}_{-l})$. Using Eqs.~(\ref{matrix-bs}) through
(\ref{matrix-total}), we solve Eq.~(\ref{input-output}) to obtain
the following operator relations:
\begin{subequations}
\begin{align}
\hat{s}_{+}^{\dagger}=k_1 \hat{a}_{+}^{\dagger}+ik_2
b_{-}^{\dagger}, \qquad \hat{s}_{-}^{\dagger}=k_1
\hat{a}_{-}^{\dagger}+ik_2 b_{+}^{\dagger}; \\
\hat{i}_{+}^{\dagger}=ik_4 \hat{a}_{-}^{\dagger}+k_3
b_{+}^{\dagger}, \qquad \hat{i}_{-}^{\dagger}=ik_4
\hat{a}_{+}^{\dagger}+k_3 b_{-}^{\dagger};
\end{align}\label{opertor-relation}
\end{subequations}
where $k_1=k_3^{*}=\frac{1}{2}(-1-e^{2il\theta})$,
$k_2=k_4^{*}=\frac{1}{2}(-1+e^{2il\theta})$. With the above
operator relations, we next calculate the angular resolution and
sensitivity that can be obtained with two and four entangled
photons.

\subsection{Supersensitive measurement with two entangled photons}

\begin{figure}
\centering
\includegraphics{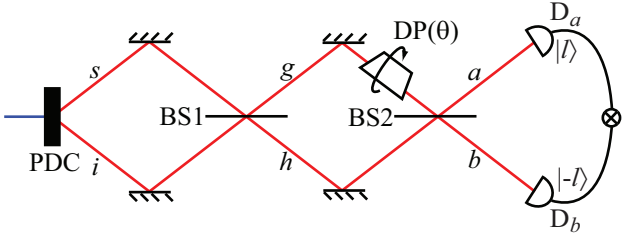}
\caption{ Scheme for supersensitive measurement of angular
displacement using two entangled photons. $\theta$ is the angle of
rotation of the Dove prism in mode $g$. $D_{a}$ and $D_b$ are
detectors set to detect one photon each in the mode characterized
by index $l$ and $-l$, respectively}. \label{fig1}
\end{figure}
In this subsection, we illustrate supersensitive
angular-displacement measurements with two entangled photons using
the detection scheme depicted in Fig.~\ref{fig1}. We consider a
class of states $|\psi_2^l\rangle$ that are obtained from
Eq.~(\ref{two-photon-state1}) by keeping only terms with the OAM
mode indices $\pm l$ for a given value of $l$:
\begin{align}
|\psi_2^l\rangle
=\sqrt{\frac{1}{2}}\left[|1\rangle_{s_{+l}}|1\rangle_{i_{-l}}+|1\rangle_{s_{-l}}|1\rangle_{i_{+l}}
\right]. \label{two-photon-state2}
\end{align}
In practice, such states can be obtained from the state produced
by PDC by placing appropriate phase apertures in the paths of the
signal and idler photons \cite{leach2009optexp}. We note that we
have used the alternative notation in writing the state
$|\psi_2^l\rangle$, which can be written in terms of the input
mode creation operators as
\begin{align}
|\psi_2^l\rangle =\sqrt{\frac{1}{2}}\left[\hat{s}^{\dagger}_{+l}
\hat{i}^{\dagger}_{-l}+\hat{s}^{\dagger}_{-l}
\hat{i}^{\dagger}_{+l} \right]|{\rm vac}\rangle.
\end{align}
Using the operator relations of Eq.~(\ref{opertor-relation}), we
express the above state in terms of the output mode creation
operators to obtain
\begin{multline}|
\psi_2^l\rangle =\sqrt{\frac{1}{2}}[(k_1
\hat{a}_{+l}^{\dagger}+ik_2 b_{-l}^{\dagger})(ik_4
\hat{a}_{+l}^{\dagger}+k_3 b_{-l}^{\dagger})
\\ +(k_1 \hat{a}_{-l}^{\dagger}+ik_2 b_{+l}^{\dagger})(ik_4
\hat{a}_{-l}^{\dagger}+k_3 b_{+l}^{\dagger})]|{\rm vac}\rangle.
\label{two-photon-state3}
\end{multline}

We now estimate the angular resolution and sensitivity through use
of the following measurement operator \cite{kolkiran2007optexp}:
\begin{align}
\hat{A}_2=|1\rangle_{a_{+l}}|1\rangle_{b_{-l}}{}_{a_{+l}}\langle
1|{}_{b_{-l}}\langle -1|,
\end{align}
which measures the probability of detecting a photon in mode $a$
with the OAM mode index $l$ and another photon in mode $b$ with
the OAM mode index $-l$. The measurement operator $A_2$ does not
see the complete state $|\psi_2^l\rangle$; the effective
post-selected state $|\psi_2^l\rangle_{\rm post}$ that $\hat{A}_2$
sees is obtained from $|\psi_2^l\rangle$ by keeping only the terms
containing $\hat{a}_{+l}^{\dagger}\hat{b}_{-l}^{\dagger}$.
$|\psi_2^l\rangle_{\rm post}$ is given by:
\begin{align}
|\psi_2^l\rangle_{\rm
post}&=\frac{1}{\sqrt{2}}(k_1^2-k_2^2)\hat{a}_{+l}\hat{b}_{-l}|{\rm
vac}\rangle \notag
\\
&=-\frac{1}{2\sqrt{2}}\left[1+e^{4il\theta}\right]|1\rangle_{a_{+l}}|1\rangle_{b_{-l}}.
\end{align}
Taking the inner product of the above state with itself, we obtain
${}_{\rm post}\langle \psi_2^l | \psi_2^l\rangle_{\rm post}=
\frac{1}{2}\cos^22l\theta$. We note that even in the best case in
which $\theta=0$, only half of the input state is detected in
modes $a_{+l}$ and $b_{-l}$, since the other half of the state
ends up in modes $a_{-l}$ and $b_{+l}$. The expectation value of
the measurement operator $\hat{A}_2$ is
$\langle\hat{A}_2\rangle={\rm
Tr}[\hat{A}_2|\psi_2^l\rangle\langle\psi_2^l|]={\rm
Tr}[\hat{A}_2|\psi_2^l\rangle_{\rm post}{}_{\rm
post}\langle\psi_2^l|]=\cos^2 (2 l\theta)$. We thus see that there
is a two-fold enhancement in the resolution of angular
displacement measurements. Next, noting that
$\langle\hat{A}_2^2\rangle=\langle\hat{A}_2\rangle=\cos^2 (2
l\theta)$, we obtain
\begin{align}
\langle\Delta\hat{A}_2\rangle=\sqrt{\langle\hat{A}_2^2\rangle-\langle\hat{A}_2\rangle^2}=\frac{\sin(4l\theta)}{2}.
\end{align}
Therefore, the uncertainty $\Delta\theta$ in the estimation of the
angular displacement is
\begin{align}
\Delta\theta=\frac{\langle\Delta\hat{A}_2\rangle}
{|\partial\langle\hat{A}_2\rangle\big/\partial\theta|}=\frac{1}{4l}.
\end{align}
In the next subsection we calculate the angular resolution and
sensitivity using four entangled photons and also with $N$
entangled photons. We then compare these results with the angular
resolution and sensitivity obtained with $N$ one-photon Fock state
input.

\subsection{Angular super-sensitivity with four entangled photons}

\begin{figure}
\centering
\includegraphics{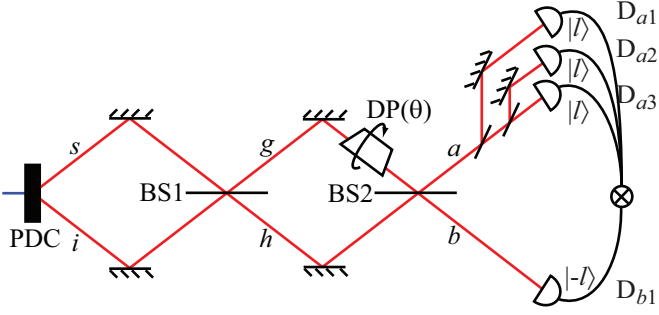}
\caption{A scheme for supersensitive measurement of angular
displacement using four entangled photons. $\theta$ is the angle
of rotation of the Dove prism in mode $g$. $D_{a1}$, $D_{a2}$,
$D_{a3}$ and $D_{b1}$ are detectors set to detect one photon each
in the mode characterized by index $l$, $l$, $l$, and $-l$,
respectively.} \label{fig2}
\end{figure}

We now consider the class of four-photon entangled states
$|\psi_4^l\rangle$ that is obtained from
Eq.~(\ref{four-photon-state1}) by keeping only terms with OAM mode
indices $\pm l$ for a given value of $l$, that is
\begin{multline}
|\psi_4^l\rangle
=\frac{1}{2}\Big[|2\rangle_{s_{+l}}|2\rangle_{i_{-l}}
+|2\rangle_{s_{-l}}|2\rangle_{i_{+l}} + \\
|1\rangle_{s_{+l}}|1\rangle_{s_{-l}}|1\rangle_{i_{+l}}|1\rangle_{i_{-l}}
+
|1\rangle_{s_{-l}}|1\rangle_{s_{+l}}|1\rangle_{i_{-l}}|1\rangle_{i_{+l}}
\Big]. \label{four-photon-state2}
\end{multline}
The state $|\psi_4^l\rangle$ can be written in terms of the input
mode creation operators as
\begin{multline}
|\psi_4^l\rangle
=\frac{1}{2}[\hat{s}^{\dagger}_{+l}\hat{s}^{\dagger}_{+l}\hat{i}^{\dagger}_{-l}\hat{i}^{\dagger}_{-l}
+\hat{s}^{\dagger}_{-l}\hat{s}^{\dagger}_{-l}\hat{i}^{\dagger}_{+l}\hat{i}^{\dagger}_{+l}\\
+
2\hat{s}^{\dagger}_{+l}\hat{s}^{\dagger}_{-l}\hat{i}^{\dagger}_{-l}\hat{i}^{\dagger}_{+l}]|{\rm
vac}\rangle.
\end{multline}
Our four-photon measurement operator $\hat{A}_4$ is
\begin{align}
\hat{A}_4=|3\rangle_{a_{+l}}|1\rangle_{b_{-l}}{}_{a_{+l}}\langle
3|{}_{b_{-l}}\langle 1|,
\end{align}
which measures the probability of detecting three photons in mode
$a_{+l}$ and one photon in mode $b_{-l}$. In Fig.~\ref{fig2}, we
have depicted one possible way of carrying out such a measurement
\cite{kolkiran2007optexp}. The particular choice of the
measurement operator is motivated by the fact that in order to
achieve super-sensitivity, the four-photon measurement needs to
post-select the ensemble that consists only of the maximally
entangled four-photon states. The effective post-selected state
$|\psi_4^l\rangle_{\rm post}$ that the measurement operator
$\hat{A}_4$ sees is obtained by first expressing the state
$|\psi_4^l\rangle$ in terms of the output state creation operators
using the operator relations given in Eq.~(\ref{opertor-relation})
and then keeping only the terms containing
$\hat{a}^{\dagger}_{+l}\hat{a}^{\dagger}_{+l}\hat{a}^{\dagger}_{+l}\hat{b}^{\dagger}_{-l}$
in the expansion. After a straightforward calculation we obtain:
\begin{align}
|\psi_4^l\rangle_{\rm post} &=\frac{1}{4}(-2ik_1^3k_2-2ik_1k_2^3)
\hat{a}^{\dagger}_{+l}\hat{a}^{\dagger}_{+l}\hat{a}^{\dagger}_{+l}\hat{b}^{\dagger}_{-l}|{\rm
vac}\rangle \notag \\
&=\frac{-i\sqrt{6}}{16}\left(1-e^{8il\theta}\right)|3\rangle_{a_{+l}}|1\rangle_{b_{-l}}.
\end{align}
Now taking the inner product of the state with itself, we find
${}_{\rm post}\langle \psi_4^l | \psi_4^l\rangle_{\rm post}=
\frac{3}{32}\sin^24l\theta$. We note that even in the best case,
when $4l\theta=\pi/2$, only 3/32nd of the input intensity is
detected by the measurement operator. The expectation value of the
measurement operator is
$\langle\hat{A}_4\rangle=\langle\hat{A}_4^2\rangle={\rm
Tr}[\hat{A}_4|\psi_4^l\rangle\langle\psi_4^l|]={\rm
Tr}[\hat{A}_4|\psi_4^l\rangle_{\rm post}{}_{\rm
post}\langle\psi_4^l|]=\sin^2 (4l\theta)$, thus showing a
four-fold enhancement in angular resolution. The uncertainty
$\langle\Delta\hat{A}_4\rangle$ is given by
$\langle\Delta\hat{A}_4\rangle=\sqrt{\langle\hat{A}_4\rangle-\langle\hat{A}_4\rangle^2}=\sin(8l\theta)/2
$. Therefore, for the uncertainty $\Delta\theta$ in the estimation
of angular displacement, we obtain
\begin{align}
\Delta\theta=\frac{\langle\Delta\hat{A}_4\rangle}
{|\partial\langle\hat{A}_4\rangle\big/\partial\theta|}=\frac{1}{8l}.
\end{align}

\subsection{Angular super-sensitivity with $N$ entangled photons}

Proceeding in the same manner, one can show that, with $N$
entangled photons and a suitable detection scheme, the angular
resolution gets enhanced by a factor of $N$ and the angular
sensitivity improves as
\begin{align}
\Delta\theta=\frac{1}{2Nl}. \label{angular-sensitivity}
\end{align}
We note that the angular sensitivity obtained with $N$ entangled
photons is greater than the angular sensitivity $\Delta\theta_{\rm
Fock}$ obtained with a stream of $N$ one-photon Fock state input
to an interferometer, which can be shown to be $\Delta\theta_{\rm
Fock}=1/(2\sqrt{N}l)$ \cite{scully1993pra}. We further note that
Eq.~(\ref{angular-sensitivity}) is the maximum angular sensitivity
that can be obtained with $N$ entangled photons. In deriving
Eq.~(\ref{angular-sensitivity}), we have not included other
factors, such as the effects due to post-selection, that also
affect the sensitivity in a laboratory situation. Some of these
factors have been discussed by Okamoto et al.
\cite{okamoto2008njp} in their work on phase super-sensitivity.

\section{Conclusions}

In conclusion, we have shown that the use of entangled photons
having non-zero orbital angular momentum increases the resolution
and sensitivity of angular-displacement measurements performed
using an interferometer. Using a $4\times4$ matrix formulation to
study the propagation of entangled OAM modes, we have explicitly
analyzed measurement schemes for two and four entangled photons.
We have found that the resolution of angular-displacement
measurements scales as $Nl$ while the angular sensitivity
increases as $1/(2Nl)$, where $N$ is the number of entangled
photons and $l$ the magnitude of the orbital-angular-momentum mode
index. It has previously been established \cite{mair2001nature,
leach2010science} that the orbital angular momentum of light
constitutes a useful degree of freedom for applications in quantum
optics and quantum information science. The work presented here
provides another such example. The ability to detect small
rotations of optical components or of light beams themselves holds
promise for many applications both in remote sensing and for
performing fundamental studies of the propagation of light through
optical materials.

\section*{ACKNOWLEDGMENTS}

We gratefully acknowledge financial support through a MURI grant
from the U.S. Army Research Office and by the DARPA InPho program
through the US Army Research Office award W911NF-10-1-0395.

%\bibliographystyle{prsty}
%\bibliography{thesis}

\end{document}